 \documentclass[final,authoryear,5p,times,twocolumn]{elsarticle}

\usepackage{graphicx}

\usepackage{amssymb}

\journal{Planetary and Space Science}

\begin{document}

\begin{frontmatter}



\title{Dense molecular cloud cores as a source of
micrometer-sized grains in galaxies}


\author[label1]{Hiroyuki Hirashita}
\ead{hirashita@asiaa.sinica.edu.tw}
\author[label2]{Ryosuke S. Asano}
\author[label3]{Takaya Nozawa}
\author[label4]{Zhi-Yun Li}
\author[label1]{Ming-Chang Liu}
\address[label1]{Institute of Astronomy and Astrophysics,
Academia Sinica, P.O. Box 23-141, Taipei 10617, Taiwan}
\address[label2]{Department of Particle and Astrophysical
Science, Nagoya University, Furo-cho, Chikusa-ku, Nagoya
464-8602, Japan}
\address[label3]{Kavli Institute for the Physics and Mathematics of the Universe (WPI),
University of Tokyo, Kashiwa, Chiba 277-8583, Japan}
\address[label4]{Astronomy Department, University of Virginia,
Charlottesville, VA 22904, USA}

\begin{abstract}
Coreshine in dense molecular cloud cores (dense cores) is
interpreted as evidence for micrometer-sized grains
(referred to as very large grains, VLGs). VLGs may have a
significant influence on the total dust amount and the
extinction curve. We estimate the total abundance of VLGs
in the Galaxy, assuming that dense cores are the site of
VLG formation. We find that the VLG abundance relative
to the total dust mass is roughly
$\phi_\mathrm{VLG}\sim 0.01(1-\epsilon )/\epsilon
(\tau_\mathrm{SF}/5\times 10^9~\mathrm{yr})^{-1}
(f_\mathrm{VLG}/0.5)(t_\mathrm{shat}/10^8~\mathrm{yr})$,
where $\epsilon$ is the star formation efficiency in dense
cores, $\tau_\mathrm{SF}$ the timescale of gas consumption
by star formation, $f_\mathrm{VLG}$ the fraction of
dust mass eventually coagulated into VLGs in dense cores,
and $t_\mathrm{shat}$ the lifetime of VLGs
(determined by shattering). Adopting their typical values for the
Galaxy, we obtain
$\phi_\mathrm{VLG}\sim 0.02$--0.09. This abundance is
well below the value detected in the heliosphere by
\textit{Ulysses} and \textit{Galileo}, which means that
local enhancement of VLG abundance in the solar
neighborhood is required if the VLGs originate from
dense cores. We also show that the effects of VLGs on
the extinction curve are negligible even with the upper
value of the above range,
$\phi_\mathrm{VLG}\sim 0.09$. If we adopt an extreme
value, 
$\phi_\mathrm{VLG}\sim 0.5$, close to that inferred from
the above spacecraft data, the extinction curve is still
in the range of the variation in Galactic extinction curves,
but is not typical of the diffuse ISM.
\end{abstract}

\begin{keyword}
Dust \sep Galaxy evolution \sep Grain size distribution
\sep Milky Way \sep Interstellar medium
\end{keyword}

\end{frontmatter}


\section{Introduction}
\label{sec:intro}

Dust grains play an essential role in some fundamental
physical processes in the interstellar medium (ISM).
First, they dominate the radiative
transfer of stellar light in the ISM. In particular,
the extinction curve, that is, the wavelength
dependence of optical depth for dust absorption plus scattering
is known to reflect the dust materials
\citep[e.g.,][]{hoyle69} and grain size distribution
(e.g., \citealt{mathis77}, hereafter MRN; \citealt{draine03}).
Second, the dust surface is the main site for the formation of some
molecular species, especially H$_2$
\citep[e.g.,][]{gould63,cazaux02}.
The rate of dust surface reaction
is proportional to the total
surface area of dust grains
\citep[e.g.,][]{hollenbach71,yamasawa11}.
Since the extinction curve and the total grain surface area both
depend strongly on the grain size distribution, clarifying the
regulating mechanism of grain size distribution is of particular
importance in understanding those important roles of dust in the
ISM.

MRN {showed} that a mixture of silicate and graphite with
a grain size distribution (number of grains per grain radius)
proportional to $a^{-3.5}$, where $a$ is the grain radius
($a=0.001$--$0.25~\mu$m), reproduces the Milky Way extinction curve.
This size distribution is {referred to as the MRN size distribution}.
\citet{kim94} and
\citet{weingartner01} {made more detailed models of}
the Milky Way extinction curve. In {both of these models},
the abundance of
grains whose radii are beyond the maximum in the MRN size distribution
(0.25~$\mu$m) is so low that the contribution of such large grains
to the total dust mass is negligible.

{The existence of micrometer ($\mu$m)-sized grains is} suggested
in dense molecular cloud cores (called dense cores in this paper).
The so-called ``coreshine'' refers to emission in the mid-infrared
(especially the 3.6-$\mu$m Spitzer Infrared Array Camera (IRAC) band)
from deep inside dense cores of molecular clouds
\citep{steinacker10,pagani10}. It is {detected in about half
of the cores studied by \citet{pagani10}}. The emission is
interpreted as light scattered by dust grains with typical sizes of
$\sim 1~\mu$m, which is much larger than the maximum grain radius in
the diffuse interstellar medium ($\sim 0.25~\mu$m; MRN). We
{refer to $\mu$m-sized grains as}
``very large grains (VLGs)'' in this paper.

{Formation of VLGs by coagulation in dense cores has
been theoretically investigated by \citet{hirashita13}
\citep[see also][]{ormel09,ormel11}. Based on the timescale on which
grains grow up to $\mu$m sizes by coagulation, they
argued that dense cores are sustained over several free-fall
times. Since their main aim was to constrain the lifetime of
dense cores, the impact of VLG formation on the grain size
distribution in the entire ISM was beyond their scope.}
A certain fraction of VLGs formed in dense cores may be
{injected} into the diffuse ISM when the dense cores
disperse. Efficient formation of VLGs would contradict
the MRN grain size distribution in which the maximum grain
radius is $\sim 0.25~\mu$m. Thus,
based on an estimation of
the total abundance of VLGs in the Milky Way, we
examine the consistency between the formation of VLGs
suggested by
coreshine and the Galactic extinction curve
{(or the MRN grain size distribution)}. In this paper, the
abundance of VLGs stands for the ratio of the total VLG
mass to the total dust mass (including VLGs), and is denoted
as $\phi_\mathrm{VLG}$ (see Section \ref{sec:estimate}).
By definition, $0\leq\phi_\mathrm{VLG}\leq 1$.

{There are some indications that VLGs exist in the ISM.
One of the indications of interstellar VLGs is provided by meteorites.}
Large interstellar grains ($> 1~\mu$m) {are} known to exist
in chondritic meteorites. Such grains were identified based on
their extremely anomalous (way off from the average solar system
isotope ratios) isotopic compositions \citep{clayton04}.
This implies that interstellar dust grains must have resided and
survived in a dense core that ended up forming
the solar system. 

{Another indication of interstellar VLGs comes from
direct detection of interstellar grains in the heliosphere by
\textit{Ulysses} and \textit{Galileo}. These experiments have shown
that the volume mass density of VLGs is comparable to the total
dust volume mass density} derived from the typical dust-to-gas ratio
in the diffuse ISM in the Galaxy \citep{landgraf00,kruger07,frisch13}.
This seems contradictory {to} the above grain size distribution
derived by MRN, who {found} that most of the dust grains
have radii less than 0.25 $\mu$m. Thus, it has been argued
that the density of VLGs is enhanced in the solar neighborhood
\citep{draine09,frisch13}. Nevertheless, it is still
interesting to compare the
VLG abundance expected from the formation in dense cores
with the measurements, in order to quantify what fraction
of the observed VLGs can be explained by the formation
in dense cores.

We may also need to consider stellar sources of dust grains,
especially asymptotic giant branch
(AGB) stars and supernovae (SNe) for the production of VLGs.
Indeed, the size distribution of grains produced
by AGB stars is suggested to be biased toward large
($\gtrsim 0.1~\mu$m) sizes from the observations of spectral
energy distributions
{\citep{groenewegen97,gauger99,norris12}},
although \citet{hofmann01} showed that the grains are not
single-sized. Theoretical studies have also shown that the
dust grains formed in the winds of AGB stars have
typical sizes $\gtrsim 0.1~\mu$m
\citep{winters97,yasuda12}. SNe (Type II SNe) are
also considered to produce relatively large
($>0.01~\mu$m) grains because small grains are
destroyed by reverse shocks before they are ejected into
the ISM \citep{nozawa07,bianchi07}.
{However, the timescale of dust supply from stars
is longer than the shattering timescale
\citep{hirashita10} by an order of magnitude.}
Therefore, even if VLGs are supplied from stars,
they probably fail to survive in the ISM.
In this paper, we do not treat the stellar
production of VLGs because of the difficulty in
their survival, but focus on their formation in
dense cores, motivated by the new
evidence of VLGs -- coreshine.

In this paper, we estimate the abundance of VLGs in the Galaxy,
assuming that dense cores are the main sites for the formation
of VLGs. This paper is organized as follows.
In Section \ref{sec:estimate}, we formulate and estimate the
abundance of VLGs in the Galaxy. In Section \ref{sec:obs},
we compare our estimates with some observations.
In Section \ref{sec:discussion}, we discuss our results
and implications for the dust evolution in galaxies.
In Section \ref{sec:conclusion}, we give our conclusions.

\section{Estimate of the total VLG mass}\label{sec:estimate}

\subsection{Formation rate of VLGs in the Galaxy}

We estimate the supply rate of VLGs ($\mu$m-sized grains)
in the Galaxy. Motivated by coreshine as evidence of VLGs in
dense cores,
we examine the hypothesis that dense cores are the main
sites for the formation of VLGs in the Galaxy (see also
Introduction).
We assume that all dense molecular cloud
cores (dense cores) eventually convert
a significant fraction of dust grains into VLGs
by coagulation after their lifetimes. The formation rate of
VLGs
{(the total mass of VLGs is denoted as $M_\mathrm{VLG}$)}
in dense cores in the Galaxy, $[dM_\mathrm{VLG}/dt]_\mathrm{form}$,
is estimated as
\begin{eqnarray}
\left[\frac{dM_\mathrm{VLG}}{dt}\right]_\mathrm{form}\equiv
\frac{X_\mathrm{core}M_\mathrm{dust}(1-\phi_\mathrm{VLG})f_\mathrm{VLG}(1-\epsilon )}
{\tau_\mathrm{core}},
\label{eq:dmdt}
\end{eqnarray}
where $X_\mathrm{core}$ is the mass fraction of dense cores to
the total gas mass, $M_\mathrm{dust}$ is the total dust mass in
the Galaxy
($X_\mathrm{core}M_\mathrm{dust}$ is the total dust mass contained
in the dense cores),
$\phi_\mathrm{VLG}\equiv M_\mathrm{VLG}/M_\mathrm{dust}$ is the
ratio of the VLG mass to the total dust mass (the
factor $1-\phi_\mathrm{VLG}$
means that we need to subtract
the dust that has already become VLGs),
$f_\mathrm{VLG}$ is the fraction of dust that is eventually
coagulated to $\mu$m sizes in the dense cores, $\epsilon$
is the star formation
efficiency in the dense cores (the {factor} $1-\epsilon$ means that
the gas that is not included in
stars is assumed to be dispersed into the ISM), and
$\tau_\mathrm{core}$ is the lifetime of dense core (i.e., the
timescale of VLG formation).
{Note that Eq.\ (\ref{eq:dmdt}) \textit{should not} be regarded as
an ordinary differential equation, but just gives an
estimate for the VLG formation rate.}
Since dense cores are also the sites of
star formation, the star formation rate of the Galaxy is
estimated by dividing the total gas mass contained in the dense cores
with their lifetime (i.e., the timescale of star formation):
\begin{eqnarray}
\psi =\frac{\epsilon X_\mathrm{core}M_\mathrm{gas}}{\tau_\mathrm{core}},
\label{eq:sfr}
\end{eqnarray}
where $M_\mathrm{gas}$ is the total gas mass in the Galaxy
($X_\mathrm{core}M_\mathrm{gas}$ is the total gas mass in
dense cores).
This equation converts the core formation rate
($X_\mathrm{core}M_\mathrm{gas}/\tau_\mathrm{core}$) to
the star formation rate, and serves to eliminate the
core formation rate, which is unknown observationally
compared with the star formation rate.
By introducing the dust-to-gas ratio,
$\mathcal{D}\equiv M_\mathrm{dust}/M_\mathrm{gas}$
and using Eq.\ (\ref{eq:sfr}), we obtain
\begin{eqnarray}
\frac{X_\mathrm{core}}{\tau_\mathrm{core}}=
\frac{\mathcal{D}\psi}{\epsilon M_\mathrm{dust}}.\label{eq:core}
\end{eqnarray}
Inserting Eq.\ (\ref{eq:core}) into Eq.\ (\ref{eq:dmdt}), we finally
get the following estimate for the VLG formation rate:
\begin{eqnarray}
\left[\frac{dM_\mathrm{VLG}}{dt}\right]_\mathrm{form}=
\frac{1-\epsilon}{\epsilon}\,
\mathcal{D}(1-\phi_\mathrm{VLG})\psi f_\mathrm{VLG}.
\label{eq:dmdt2}
\end{eqnarray}
{This VLG formation rate} could be implemented in
a larger framework of
dust enrichment, which is capable of calculating the evolution
of the total dust mass in the Galaxy, to calculate the evolution
of $M_\mathrm{VLG}$ in a consistent way with $M_\mathrm{dust}$
{or $\mathcal{D}$}.
However, this is not necessary for the purpose of estimating the
total VLG mass in the Galaxy. The timescale
of dust enrichment
{(i.e., the timescale of the variation of $M_\mathrm{dust}$
or $\mathcal{D}$)}
in the Galaxy is roughly the metal-enrichment timescale
($\sim $ several Gyr) \citep[e.g.][]{dwek98,zhukovska08,inoue11,asano13a},
which is much longer than the lifetime of VLGs
(typically determined by the shattering timescale
{$\sim 10^8$ yr; \citealt{hirashita10}).
Therefore, we can assume that $M_\mathrm{dust}$ and $\mathcal{D}$
are constant within the lifetime of VLGs.}
In such a case, the total mass of VLGs can be approximately
estimated as follows.

It is shown that $\mu$m-sized grains are shattered
in the diffuse ISM by grain--grain collisions under
the grain motion induced by turbulence
\citep{yan04,hirashita10}. Shattering also occurs
in supernova shocks \citep{jones96}. Thus, we assume that
the lifetime of VLGs is determined by the shattering
timescale, $t_\mathrm{shat}$ ($\sim 10^8$ yr; \citealt{hirashita10}).
{The destruction rate of VLGs can thus be approximately
estimated as $M_\mathrm{VLG}/t_\mathrm{shat}$, and
the equilibrium between the VLG formation and destruction
is achieved on a timescale of $\sim t_\mathrm{shat}$.
Since, as mentioned above, $t_\mathrm{shat}$ is much
shorter than the evolution timescale of the dust mass,
the variation of $\mathcal{D}$ in $t_\mathrm{shat}$ can
be neglected. Therefore, the
total mass of VLGs in the Galaxy is estimated by
the equilibrium between the formation and destruction
under a fixed $\mathcal{D}$:}
\begin{eqnarray}
M_\mathrm{VLG} & \sim &
\left[\frac{dM_\mathrm{VLG}}{dt}\right]_\mathrm{form}
\cdot t_\mathrm{shat}\nonumber\\
& = & \frac{1-\epsilon}{\epsilon}\,
\mathcal{D}(1-\phi_\mathrm{VLG})\psi f_\mathrm{VLG}t_\mathrm{shat}.
\label{eq:M_VLG_main}
\end{eqnarray}

\subsection{Estimates of various quantities}\label{subsec:estimate}

We adopt dust-to-gas ratio $\mathcal{D}=0.01$,
the same value as adopted
in our previous calculations of coagulation in dense cores
\citep{hirashita13}. The star formation rate in the Galaxy
can be estimated from the total luminosity of OB stars.
\citet{hirashita07} obtained a star formation rate of
1.3 M$_\odot~\mathrm{yr}^{-1}$ based on the total OB star luminosity
derived observationally by \citet{mathis83}. According to
\citet{hirashita13},
almost all the dust mass is in VLGs
after coagulation in dense cores. Here we conservatively
assume that half of the dust mass is converted to
VLGs after coagulation in dense cores (i.e., $f_\mathrm{VLG}=0.5$).
For $t_\mathrm{shat}$, we adopt $10^8$ yr according to
\citet{hirashita10},
{who considered shattering under the grain motion
driven by interstellar
turbulence. As calculated by \citet{jones96}, shattering
can also take place in supernova shocks.} Even if the
supernova shocks are
the main site of shattering
\citep{jones96}, a similar timescale is
obtained for shattering (Section \ref{subsec:variation}).

Using the above values, we obtain
\begin{eqnarray}
M_\mathrm{VLG} & \sim & 5\times 10^5\,(1-\phi_\mathrm{VLG})\,
\frac{1-\epsilon}{\epsilon}
\left(\frac{\mathcal{D}}{0.01}\right)\nonumber\\
& &
\times\left(\frac{\mathcal{\psi}}{1~\mathrm{M}_\odot~\mathrm{yr}^{-1}}\right)
\left(\frac{f_\mathrm{VLG}}{0.5}\right)
\left(\frac{t_\mathrm{shat}}{10^8~\mathrm{yr}}\right)~\mathrm{M}_\odot .
\label{eq:M_VLG}
\end{eqnarray}
For comparison, we also estimate the total dust mass.
The total gas mass in the Milky Way is
$M_\mathrm{gas}\sim 5\times 10^9$ M$_\odot$
\citep{mathis00,tielens05}. By multiplying the dust-to-gas
ratio, the dust mass is estimated as
\begin{eqnarray}
M_\mathrm{dust}=M_\mathrm{gas}\mathcal{D}\sim 5\times 10^7
\left(\frac{\mathcal{D}}{0.01}\right)
\left(\frac{M_\mathrm{gas}}{5\times 10^9~\mathrm{M}_\odot}\right)
~\mathrm{M}_\odot .\label{eq:M_dust}
\end{eqnarray}
Dividing Eq.\ (\ref{eq:M_VLG}) with Eq.\ (\ref{eq:M_dust}),
we obtain the following expression by recalling that
$\phi_\mathrm{VLG}=M_\mathrm{VLG}/M_\mathrm{dust}$:
\begin{eqnarray}
\frac{\phi_\mathrm{VLG}}{1-\phi_\mathrm{VLG}}=
0.01\,\frac{1-\epsilon}{\epsilon}\,
\left(\frac{\tau_\mathrm{SF}}{5\times 10^9~\mathrm{yr}}\right)^{-1}
\left(\frac{f_\mathrm{VLG}}{0.5}\right)
\left(\frac{t_\mathrm{shat}}{10^8~\mathrm{yr}}\right) ,
\label{eq:phi}
\end{eqnarray}
where $\tau_\mathrm{SF}\equiv M_\mathrm{gas}/\psi$ is the
star formation (gas consumption) time.
The star formation efficiency $\epsilon$ in molecular cloud
cores is around 0.1--0.3 \citep{alves07,curtis10,lada10}.
Thus, we can assume that $\phi_\mathrm{VLG}\ll 1$ in the
Galaxy. In this case, we simply replace the left-hand side
of Eq.\ (\ref{eq:phi}) with
$\phi_\mathrm{VLG}$.

\section{Comparison with observational data}\label{sec:obs}

\subsection{Direct detection}

\begin{figure}[ht]
\begin{center}
\includegraphics[width=0.48\textwidth]{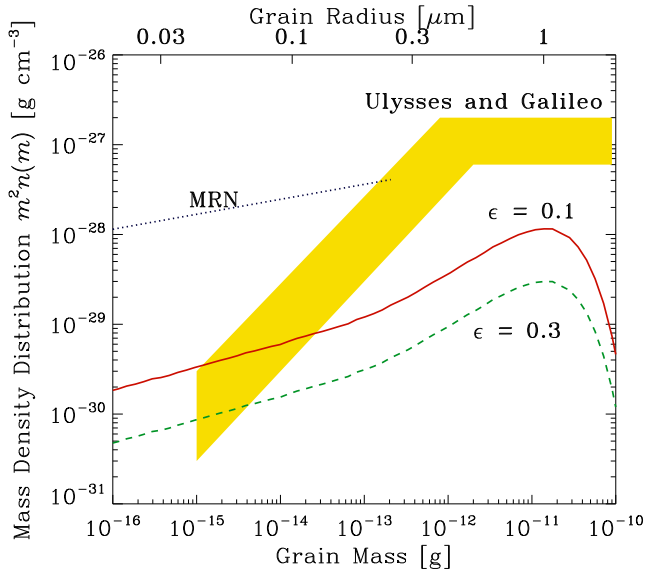}
\end{center}
\caption{
Grain mass distributions. The solid and dashed lines
show the grain mass distributions calculated by
\citet{hirashita13}, but scaled with the total
VLG abundance appropriate for $\epsilon =0.1$ and 0.3,
respectively. The dotted line shows the MRN distribution,
which is considered to be representative of the grain
size distribution in the diffuse ISM. The data taken by
the \textit{Ulysses} and \textit{Galileo} spacecraft are
represented by the shaded area, the width of which
shows the typical error \citep{frisch13}.
The corresponding grain radii are also shown on
the upper axis with grain material density 3.3 g cm$^{-3}$.
}\label{fig:distribution}
\end{figure}

Interstellar dust grains with {radii}
$a\gtrsim 0.1~\mu$m can be
detected directly {in space} \citep{mann09}.
In Fig.\ \ref{fig:distribution}, we show
the grain mass distribution for interstellar grains
in the heliosphere observed by
the \textit{Ulysses} and \textit{Galileo} spacecraft
\citep{frisch13}. {Small
grains with typically $a<0.1~\mu$m are excluded
from heliospheric
plasma because of large charge-to-mass ratios. Thus,}
we are only interested in the data at
$a> 0.1~\mu$m.

In Fig.\ \ref{fig:distribution}, we
also plot the grain mass distribution calculated by
\citet{hirashita13} but scaled so that
the total mass density of VLGs is
$\phi_\mathrm{VLG}\mathcal{D}\mu m_\mathrm{H}n_\mathrm{H}$,
where $\mu$ is the gas mass per hydrogen atom
(1.4), $m_\mathrm{H}$ is the mass of hydrogen atom,
$n_\mathrm{H}=0.1~\mathrm{cm}^{-3}$ is the number
density of hydrogen nuclei in the local ISM \citep{frisch13}.
\citet{draine09} adopted $n_\mathrm{H}=0.22~\mathrm{cm}^{-3}$,
but the difference by a factor of 2 does not affect the conclusions.
The functional form of the distribution is taken from
the maximal coagulation model with a number density of
hydrogen nuclei of $10^5$ cm$^{-3}$
at $t=5.5t_\mathrm{ff}$ ($t_\mathrm{ff}$ is the free-fall time),
when the peak is located at $a=1~\mu$m.
The peak roughly reflects the mass density of VLGs, and
is not sensitive to the
choice of $t$ in any case. From Eq.\ (\ref{eq:phi}),
we adopt {VLG-to-total-dust mass ratio}
$\phi_\mathrm{VLG}=0.01(1-\epsilon )/\epsilon$
with {star formation efficiency}
$\epsilon =0.1$--0.3 (Section \ref{subsec:estimate}).
Note that the grain mass distribution
is expressed as $m^2n(m)=(4\pi\rho_\mathrm{gr}/9)n'(a)a^4$,
where $\rho_\mathrm{gr}$ is
the grain material density, and $n'(a)$
is the grain size distribution,\footnote{The grain size
distribution is defined so that $n'(a)\, da$ is
the number density of dust grains with radii between
$a$ and $a+da$.} which is related to
$n(m)$ by $n(m)\, dm=n'(a)\,da$. We adopt the same
grain material density,
$\rho_\mathrm{gr}=3.3$ g cm$^{-3}$, as
{in} \citet{hirashita13}.

As we observe in Fig.\ \ref{fig:distribution},
the mass density of VLGs expected with
$\phi_\mathrm{VLG}\sim 0.02$--0.09 (for $\epsilon =0.3$--0.1)
is much lower than {observed in space}.
This means
that the large mass density observed in the
heliosphere cannot be explained by the
abundance of VLGs formed in dense cores.
{From the analysis of their velocities, the VLGs
detected in space are not likely to be the remains
of the large grains that were produced by coagulation in
the dense core inside which the Solar System formed
\citep{howe13}.}
If dense cores are the site of VLG formation,
we need some mechanism of enhancing the
VLG abundance locally to explain the
spacecraft data. The physical mechanism of
such an enhancement is still unknown
\citep{draine09}.

For comparison, we also show the
MRN \citep{mathis77} grain size distribution
($n'(a)\propto a^{-3.5}$, with the upper and lower
limits of grain radius,
0.001 $\mu$m and 0.25 $\mu$m, respectively).
The normalization of the MRN grain size distribution
is determined so that the total dust mass density is
$\mathcal{D}\mu m_\mathrm{H}n_\mathrm{H}$
with {dust-to-gas ratio} $\mathcal{D}=0.01$.
This MRN size distribution is representative of the
grain size distribution in the diffuse ISM of the
Galaxy.
The abundance of VLGs expected from
{star formation efficiency} $\epsilon =0.1$--0.3
is well below the extrapolation of the MRN grain size
distribution to $a\sim 1~\mu$m. This is consistent
with {a small VLG-to-total-dust mass ratio,
$\phi_\mathrm{VLG}\ll 1$; that is, VLGs make only a small
contribution to} the total dust mass in the Galaxy.

\subsection{Extinction curves}\label{subsec:extinc}

The wavelength dependence of interstellar extinction,
the so-called extinction curve, is a viable tool to
derive the grain size distribution. {The
MRN size distribution was derived by fitting}
the averaged interstellar extinction curve in the
Milky Way. For the MRN size distribution, the upper
limit of {the} grain radius is 0.25 $\mu$m. Here we examine
if {the inclusion of VLGs is}
consistent with the Galactic extinction curve or not.

{A possible caveat of extinction curve fitting
is that the solution is not unique. For example,
MRN adopted graphite for the carbonaceous material
because of the strong 2175 \AA\ bump, while
\citet{jones12} proposed hydrogenated amorphous carbon.
\citet{compiegne11} used the latter species for the
carbonaceous dust component and reproduced both
the extinction and emission of dust in the Galaxy.
The relative contribution between silicate and
carbonaceous dust also depends on the material
properties adopted.
Therefore, strictly speaking, adopting a specific model
reproducing the extinction curve means
that the following result only
serves as an example of the effects of VLGs.
However, our results below are valid at least qualitatively
as long as large grains tend to show flat extinction
curves in the optical and the 2175 \AA\ bump is produced
by small grains.}

We assume that grains are composed of two species:
silicate and graphite. Extinction curves are calculated
by using the same optical properties of
{silicates and graphite as in \citet{hirashita09}.
That is,} we adopt the optical constants from
\citet{draine84} and calculate extinction cross-sections
by the Mie theory \citep{bohren83}. The cross-sections
are weighted {by}
the grain size distributions, and summed up
for silicate and graphite.
The fraction of silicate to the total dust mass is
assumed to be 0.54, and the rest is {assumed} to be
graphite {\citep{takagi03,hirashita09}}.
The contributions from the MRN size
distribution and from the VLGs to the total extinction
are proportional to $(1-\phi_\mathrm{VLG})$ and
$\phi_\mathrm{VLG}$, respectively.

\begin{figure}[ht]
\begin{center}
\includegraphics[width=0.48\textwidth]{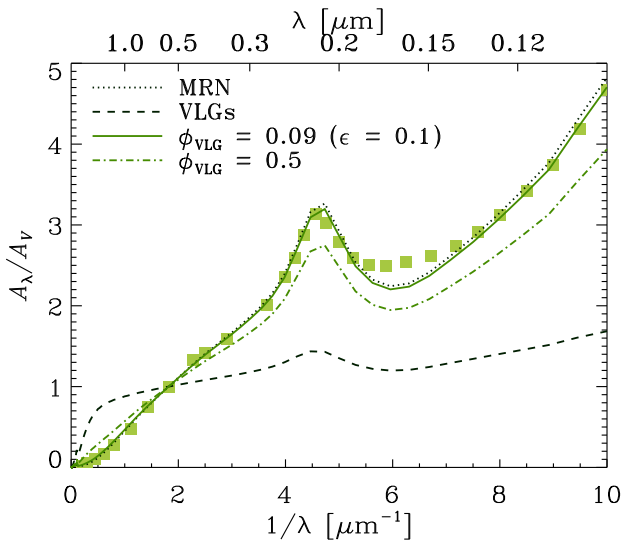}
\end{center}
\caption{
Extinction curves. The solid and dot-dashed lines
show the extinction curves calculated for
$\phi_\mathrm{VLG}=0.09$ (corresponding to
$\epsilon =0.1$) and $\phi_\mathrm{VLG}=0.5$
(as an extreme case, based on the \textit{Ulysses} and
\textit{Galileo} data),
respectively. The dotted and dashed lines show
the extinction curves for the MRN and VLG components,
respectively. The filled squares show the
observed Galactic extinction curve taken from
\citet{pei92}.
}\label{fig:extinc}
\end{figure}

In Fig.\ \ref{fig:extinc}, we show the results.
{The extinction is
normalized to the $V$ band
(0.55 $\mu$m), and the normalized extinction at
wavelength $\lambda$ is denoted as $A_\lambda /A_V$.}
We find that, even for {star formation
efficiency} $\epsilon =0.1$
(equivalent to VLG-to-total-dust mass ratio
$\phi_\mathrm{VLG}=0.09$), there is only
a slight difference from the extinction curve for
the MRN model\footnote{The extinction curve
calculated by the
MRN model has some deviations from the observational
data taken from \citet{pei92}: one is seen around
$1/\lambda\sim 6~\mu\mathrm{m}^{-1}$ and
another is the different value of $R_V$ (note that
the mean value of $R_V$ is 3.1 \citep{pei92}.
However, the overall shape of the observed mean
extinction curve (shown by filled squares in
Fig.\ \ref{fig:extinc}) is well reproduced by the
MRN grain size distribution, and further fine-tuning
does not affect the discussions and conclusion
in this paper.}
with a little enhancement in infrared
extinction, {and slightly lower carbon bump and
ultraviolet extinction}.
All these {differences}
can be explained by the contribution from VLGs,
as is clear from the extinction curve of the
VLG component (dashed line in Fig.\ \ref{fig:extinc}).
However, the difference is negligible
compared with the typical variation in the Milky Way
\citep{nozawa13}.
$R_V\equiv A_V/(A_V-A_B)$, which is an indicator of
the flatness of extinction curve, is 3.6 and
3.7 for MRN and $\epsilon =0.1$, respectively.
For $\epsilon =0.3$, the difference
is even smaller. Therefore, our estimate of
the VLG abundance is within the acceptable range
as far as the variation of extinction curve is
concerned.

We also show an extreme case where
$\phi_\mathrm{VLG}=0.5$, considering that
the above \textit{Ulysses} and \textit{Galileo} data
show a mass density of VLGs comparable to the
total dust mass density in the diffuse ISM
(i.e., the MRN component in Fig.\ \ref{fig:distribution}).
The deviation from
the mean Galactic extinction curve is clear in
this case. For this extinction curve, we obtain
$R_V=4.6$, which is
still in the range of the variation in the Milky Way
extinction curve, but is not typical of the
diffuse ISM \citep{cardelli89,fitzpatrick07}.
This again supports the view that the high VLG abundance
is due to local enhancement.
\citet{draine09} also calculated the extinction curve
based on \citet{weingartner01}'s model modified
for the \textit{Ulysses} and \textit{Galileo} measurements.
They obtained $R_V=5.8$, the difference from our
models being due to their higher abundance of VLGs.
They also conclude that
the large excess of $\mu$m-sized
interstellar grains
in the heliosphere is not representing the typical
diffuse medium in the Milky Way but is probably
due to local enhancement of
VLG abundance.

\section{Discussion}\label{sec:discussion}

\subsection{In the context of galaxy evolution}

The above estimate of $\phi_\mathrm{VLG}$
{(the VLG-to-total-dust mass ratio)} is applicable to
the Milky Way. Galaxies in general are expected to have
a variety in 
$\phi_\mathrm{VLG}$. The central hypothesis
in this paper is that VLGs form
in dense cores. This is probably true in solar-metallicity
environments where dust-to-gas ratio is large enough
for coagulation to occur efficiently in dense cores.
In contrast, in a galaxy whose
dust-to-gas ration is lower, coagulation in dense cores
may not {be} efficient enough to produce VLGs; in other words,
$f_\mathrm{VLG}$ is smaller if dust-to-gas ratio
(or metallicity) is
lower. Since dust-to-gas ratio
has a positive correlation with metallicity
\citep{schmidt93}, we expect
that $\phi_\mathrm{VLG}$ is smaller in lower-metallicity
galaxies.

\citet{asano13b} showed that the major source of grains
with $a>0.1~\mu$m is stars (SNe and AGB stars) in
low-metallicity (or low-dust-to-gas-ratio) environments.
Therefore, if coagulation is not efficient,
we expect that most of the VLGs (if they exist) are
of stellar origin.
{As mentioned in Introduction, the size distributions
of grains produced by AGB stars and SNe are suggested to
be biased toward large ($\gtrsim 0.1~\mu$m) sizes.}
However, there is still
an uncertainty in the size distribution of
dust grains formed in stellar ejecta.
For more quantitative
estimates of relative importance between stellar VLGs
and dense-core VLGs, we need to construct
a framework that takes into account all dust formation
and destruction mechanisms as done by \citet{asano13b}.
Since they did not include VLG formation in dense cores,
it is necessary to implement it into their framework
in the future.

The stellar origin of VLGs is worth considering further.
As mentioned in Introduction, large interstellar grains
($>1~\mu$m) exist in chondritic meteorites.
Given their large isotope anomalies that are reflective
of stellar nucleosynthesis processes \citep{clayton04},
the grains may have formed in stellar ejecta.
However, it is unlikely that grains formed in stellar
ejecta have such a high abundance as observed by the
spacecraft, since the timescale of dust supply from
stellar sources is {likely to be longer than the
destruction timescale by shattering
\citep{hirashita10} and sputtering \citep{mckee89,jones96}}.
Recently, \citet{jones11} have suggested that the
lifetime of silicate dust is comparable to the destruction
timescale, which means that the possibility of
VLG formation in stellar ejecta is still worth investigating.

The existence of VLGs can have a significant influence
on the extinction curve if there is local enhancement
of VLG abundance as indicated by the spacecraft
missions (see the case of
$\phi_\mathrm{VLG}=0.5$ in Fig.\ \ref{fig:extinc}).
Thus, if we are concerned with the local
variation of extinction curves, the
contribution from VLGs should be taken into
account. On global scales of a galaxy, in
contrast, the abundance of VLGs is not high
enough to affect the extinction curve, as long as
the values of the parameters in Eq.\ (\ref{eq:phi})
are similar to the Galactic values.

{The far-infrared spectral energy distribution is
also affected by the existence of VLGs. The equilibrium
temperature ($T_\mathrm{eq}$) of a dust grain depends
on the grain radius as $T_\mathrm{eq}\propto a^{1/(\beta +4)}$,
where $\beta (\sim 2)$ is the power-law index of the mass absorption
coefficient $\kappa_\nu$ as a function of frequency $\nu$,
$\kappa_\nu\propto\nu^\beta$ \citep{evans94}. Because of
this dependence,
VLGs tend to be colder than ``normal'' grains
with $a\lesssim 0.1~\mu$m. If the abundance of VLGs
is high, we should consider a significant contribution
from such a ``cold'' dust component.}

\subsection{Star formation efficiency in dense cores}

The star formation efficiency in dense cores,
$\epsilon$, is a key factor for
{the VLG-to-total-dust mass ratio}
$\phi_\mathrm{VLG}$.
It enters our formulation in two ways. {The first} is
through Eq.\ (\ref{eq:dmdt}), in which the
factor $(1-\epsilon )$ expresses the fraction of
dust that is not included in stars formed in dense
cores. The {second} is through Eq.\ (\ref{eq:sfr}), which
connects the core formation rate to
the star formation rate. As a result,
$\phi_\mathrm{VLG}\propto (1-\epsilon )/\epsilon$
(as long as $\phi_\mathrm{VLG}\ll 1$).
If $\epsilon$ is nearly unity, $\phi_\mathrm{VLG}\sim 0$,
which means that all VLGs once formed in dense
cores are eventually included in stars.
In contrast, if $\epsilon\lesssim 0.01$,
$\phi_\mathrm{VLG}\sim 1$.
{This corresponds to a case in which the total mass
of dense cores is large.}

{The volume density of VLGs measured by \textit{Ulysses} and
\textit{Galileo} is comparable to that of the entire dust population
expected from the dust-to-gas ratio in the
ISM, $\mathcal{D}=0.01$}
(MRN in Fig.\ \ref{fig:distribution}).
For example, $\phi =0.5$ corresponds to $\epsilon =0.01$,
indicating an extremely low star formation efficiency.
Considering that star formation efficiencies observationally
estimated for dense cores are $\sim 0.1$--0.3
(Section \ref{subsec:estimate}), the high VLG abundance is
probably due to local enhancement (see also Section \ref{subsec:extinc}).

\subsection{Variation of other parameters}\label{subsec:variation}

In galaxies, $\tau_\mathrm{SF}$ and $t_\mathrm{shat}$ may
also change. The gas consumption time $\tau_\mathrm{SF}$ ranges
from $\sim 10^8$~yr to $\sim 10^{11}$~yr for nearby galaxies
\citep{kennicutt98}. In a starburst galaxy with
$\tau_\mathrm{SF}\sim 10^8$~yr, if we adopt $\epsilon =0.1$,
$f_\mathrm{VLG}=0.5$ and $t_\mathrm{shat}=10^8$~yr,
we obtain $\phi_\mathrm{VLG}\simeq 0.8$. This large value is
due to
the enhanced formation rate of dense cores. Therefore,
the abundance of VLGs can be enhanced in starburst galaxies.

{If the typical dust size in starburst galaxies is
really large and VLGs are continuously supplied there,
it may have important consequence for the dust destruction
efficiency {by sputtering}. Since the timescale of
dust destruction by
thermal sputtering is proportional to the grain size
\citep{draine79}, the VLGs survive longer than smaller
grains. In particular, extreme starbursts at high
redshift seems to have more dust than can be produced
by stars with a dust lifetime expected for the
``normal-sized'' grains \citep{valiante11,mattsson11}. The
discrepancy is explained by grain growth through
accretion of gas-phase metals in the ISM
\citep{michalowski10,mattsson11,valiante11,kuo12}.
However, if the VLGs dominate the total dust mass,
their longer survival would make the
discrepancy smaller, requiring less grain growth than is
thought in previous studies. This is also true for nearby
starburst galaxies.}

However, in starburst galaxies, shattering may also be
efficient. Grains acquire larger velocities in
the ionized medium than in the neutral medium because
less dissipative magnetohydrodynamic nature of
the ionized medium leads to more efficient
gyroresonance
{(resonance between the magnetohyrodynamical
waves with the gyromotion of a grain)} acceleration of grains
\citep{yan04,hirashita09}. As a result,
$t_\mathrm{shat}$ is shorter in the ionized medium.
Therefore, if the ISM is highly ionized as a result of
high star formation activities in
starburst galaxies, $t_\mathrm{shat}$
may decrease, compensating the decrease of
$\tau_\mathrm{SF}$ in Eq.\ (\ref{eq:phi}).

In this paper, we have considered turbulence
as the source of grain motion for shattering.
Shattering also occurs in supernova shocks
\citep{jones96}. \citet{hirashita_etal10} showed that
shattering in supernova shocks is as efficient as
shattering in turbulence. They derived the
timescale of shattering in supernova shocks
$\sim 0.01\tau_\mathrm{SF}$, which gives the
same order of magnitude as $t_\mathrm{shat}$
above. If we adopt this expression,
the dependence of $\phi_\mathrm{VLG}$ on
$\tau_\mathrm{SF}$ cancels out (Eq.\ \ref{eq:phi}).
Therefore, if supernova shocks are the main
site of shattering, the VLG abundance is insensitive
to the star formation activity.

In summary, it is not clear if the abundance of
VLGs increases or decreases in starburst galaxies.
Nevertheless, the expression in Eq.\ (\ref{eq:phi})
is useful to understand the dependence of the VLG
abundance on star formation activity and shattering
timescale.

\section{Conclusion}\label{sec:conclusion}

We have estimated the abundance of very large grains
(VLGs), whose radii are typically $\gtrsim 1~\mu$m, in
the Galaxy.
Coreshine in dense molecular cloud cores (dense cores)
is taken as evidence for such VLGs. Assuming that
VLGs are formed in dense cores, we have estimated
the abundance of VLGs in the Galaxy.
The VLG abundance relative to the total dust mass is
estimated as
$\phi_\mathrm{VLG}\sim 0.01(1-\epsilon )/\epsilon
(\tau_\mathrm{SF}/5\times 10^9~\mathrm{yr})^{-1}
(f_\mathrm{VLG}/0.5)(t_\mathrm{shat}/10^8~\mathrm{yr})$,
where $\epsilon$ is the star formation efficiency in dense
cores, $\tau_\mathrm{SF}$ is the timescale of
gas consumption by star formation, $f_\mathrm{VLG}$ is
the fraction of
dust mass eventually coagulated into VLGs in dense
clouds, and $t_\mathrm{shat}$ is the lifetime of VLGs
(determined by shattering). Adopting typical star formation efficiencies
$\epsilon\sim 0.1$--0.3, we obtain
$\phi_\mathrm{VLG}\sim 0.02$--0.09 for the Galaxy. This abundance
is well below the value detected by \textit{Ulysses} and \textit{Galileo}.
Thus, if the VLGs originate from dense cores,
local enhancement of VLG abundance in the solar
neighborhood is necessary. We have also examined the effect of VLGs on
the extinction curve, finding that the effect is negligible
even with the upper value of the
above range, $\phi_\mathrm{VLG}\sim 0.09$. With
$\phi_\mathrm{VLG}\sim 0.5$, which is near the value
of the above spacecraft data, the extinction curve is still
in the range of the variation in Galactic extinction curves, but
is not typical of the diffuse ISM. This again supports {the idea} that
the high VLG abundance in the heliosphere is due to
local enhancement. Finally, it is worth noting that
the explicit dependence of $\phi_\mathrm{VLG}$ on
$\tau_\mathrm{SF}$, $f_\mathrm{VLG}$, and $t_\mathrm{shat}$,
can be used to estimate the VLG abundance in
galaxies as well as in the Milky Way.

\section*{Acknowledgment}

We are grateful to B. T. Draine and the anonymous referees
for helpful discussions and comments. HH is
supported from NSC grant NSC102-2119-M-001-006-MY3.
RSA is supported from the Grant-in-Aid for
JSPS Research under Grant No.\ 23-5514.
TN is supported by the Grant-in-Aid for Scientific
Research of the Japan Society for the Promotion of Science
(22684004, 23224004).
ZYL is supported in part by NSF AST-1313083 and NASA NNX10AH30G grants.




\end{document}